\documentclass[prl,aps,twocolumn,showpacs,bibnotes,superscriptaddress,epsf]{revtex4}

\usepackage{float}
\usepackage{graphicx}
\usepackage{dcolumn}
\usepackage{bm}
\usepackage{SIunits}
\usepackage{tabularx}

\begin{document}

\title{Electronic structure of FeS}
 \author{J. Miao} \author{X. H. Niu} \author{D. F. Xu} \author{Q. Yao} \author{Q. Y. Chen} \author{T. P. Ying} \affiliation{State Key Laboratory of Surface Physics,
Department of Physics, and Laboratory of Advanced Materials, Fudan University, Shanghai 200433, People's Republic of China} \author{S. Y. Li} \affiliation{State Key Laboratory of Surface Physics, Department of
Physics, and Laboratory of Advanced Materials, Fudan University, Shanghai 200433, People's Republic of China} \affiliation{Collaborative Innovation Center of Advanced Microstructures, Nanjing, 210093, China}
\author{Y. F. Fang} \author{J. C. Zhang} \affiliation{Materials Genome Institute, Shanghai University, Shanghai 200444, People's Republic of China} \author{S. Ideta} \affiliation{UVSOR Synchrotron Facility,
Institute for Molecular Science, National Institutes of Natural Science, Myodaiji, Okazaki 444-8585, Japan} \author{K. Tanaka} \affiliation{UVSOR Synchrotron Facility, Institute for Molecular Science,
Myodaiji, Okazaki 444-8585, Japan}
\author{B. P. Xie}\email{bpxie@fudan.edu.cn} \affiliation{State Key Laboratory of Surface Physics, Department of Physics, and Laboratory of Advanced Materials, Fudan University, Shanghai 200433, People's Republic of China} \affiliation{Collaborative Innovation Center of Advanced Microstructures, Nanjing, 210093, China}
\author{D. L. Feng}\email{dlfeng@fudan.edu.cn} \affiliation{State Key Laboratory of Surface Physics, Department of Physics, and Laboratory of Advanced Materials, Fudan
University, Shanghai 200433, People's Republic of China} \affiliation{Collaborative Innovation Center of Advanced Microstructures, Nanjing, 210093, China} \author{Fei Chen}\email{chf001@shu.edu.cn}
\affiliation{Materials Genome Institute, Shanghai University, Shanghai 200444, People's Republic of China}

\begin{abstract}

 Here we report the electronic structure of FeS, a recently identified iron-based superconductor. Our  high-resolution angle-resolved photoemission spectroscopy studies show two hole-like ($\alpha$ and
 $\beta$) and two electron-like ($\eta$ and $\delta$) Fermi pockets around the Brillouin zone center and corner, respectively, all of which exhibit moderate dispersion along $k_z$.
  However, a third hole-like band ($\gamma$) is not observed, which is expected around the zone center from band calculations and is common in iron-based superconductors. Since this band has the highest
  renormalization factor and is known to be the most vulnerable to defects, its absence in our data is likely due to defect scattering --- and yet superconductivity can exist without coherent
  quasiparticles in the $\gamma$ band. This may help resolve the current controversy on the superconducting gap structure of FeS.
Moreover, by comparing the $\beta$ bandwidths of various iron chalcogenides, including FeS, FeSe$_{1-x}$S$_x$, FeSe, and FeSe$_{1-x}$Te$_x$, we find that the $\beta$ bandwidth of FeS is the broadest. However, the band
renormalization factor of FeS is still quite large, when compared with the band calculations, which indicates sizable electron correlations.
This explains why the unconventional superconductivity can persist over such a broad range of isovalent substitution in
 FeSe$_{1-x}$Te$_{x}$ and FeSe$_{1-x}$S$_{x}$.

\end{abstract}


\maketitle
\section{I. Introduction}

Among all the iron-based superconductors, the iron-chalcogenide FeSe has the simplest layered structure yet extraordinarily rich physics \cite{MKWuTP,JZhaoNS}. It undergoes a structural transition around 90~K \cite{JZhaoNS}, then becomes superconducting near 8~K. Under high pressure, its superconducting temperature ($T_c$)  can be enhanced to 37~K \cite{MedvedevHP}. When FeSe was intercalated with potassium by liquid gating
\cite{CHENXH} or dosed with potassium on the surface \cite{WCHP},  the $T_c$ could reach 46~K. Most remarkably, FeSe monolayer film grown on SrTiO$_3$(001) substrate even exhibits $T_c$ as high as 65~K \cite{QKXueSTM,DLFengARPES2,ZXShenARPES}.

Isovalent substitution is a particularly effective way of tuning the properties of FeSe. The $T_c$ of FeSe$_{1-x}$Te$_x$ can reach 14~K \cite{MHFangPD}, while  that of FeSe$_{1-x}$S$_x$ is first enhanced by dilute S substitution, then decreases upon further substitution \cite{WatsonARPES}. Recently, it was found that FeS, in which Se is totally substituted by S, is still  superconducting with a $T_c$ of about 4.5~K \cite{HuangTP, PachmayrXRD}. During this process, the  nematic (orthorhombic) phase transition temperature  ($T_N$) in FeSe is suppressed, and  there is no structural or nematic transition in FeS \cite{DLFengARPES3, WatsonARPES, WangSHC, MahmoudTP}.

The presence of superconductivity in FeS is quite unexpected, since S substitution is expected to significantly reduce the electron correlations, which are crucial for the unconventional superconductivity \cite{JPHuSR}. For example, it has been shown that S substitution enhances the bandwidth, and thus reduces the correlation effects in K$_x$Fe$_{2-y}$Se$_{2-z}$S$_z$, which is not superconducting for $z\geq 1.4$ \cite{DLFengARPES7}. To address this issue, it is important to examine the electronic structure of FeS. Moreover, the gap structure of FeS is currently under debate.  Muon spin rotation ($\mu$SR) measurements favor the fully-gapped superconductivity coexisting with a low-moment magnetic state at low temperature in polycrystalline FeS \cite{KirschnerMSR, HolensteinMSR}. On the contrary,  thermal conductivity and specific heat measurements support the existence of a nodal gap \cite{LiTC2, WenSHC}, while scanning tunneling spectroscopy indicates a highly anisotropic or nodal  superconducting gap structure \cite{WenSTM}. Furthermore, a theoretical calculation predicted $d_{x^2-y^2}$  pairing symmetry in FeS \cite{WangDFT}. A thorough understanding of the electronic structure of FeS will contribute to these discussions.

\begin{figure*}[htb]
\centering
\includegraphics[width=0.95\textwidth]{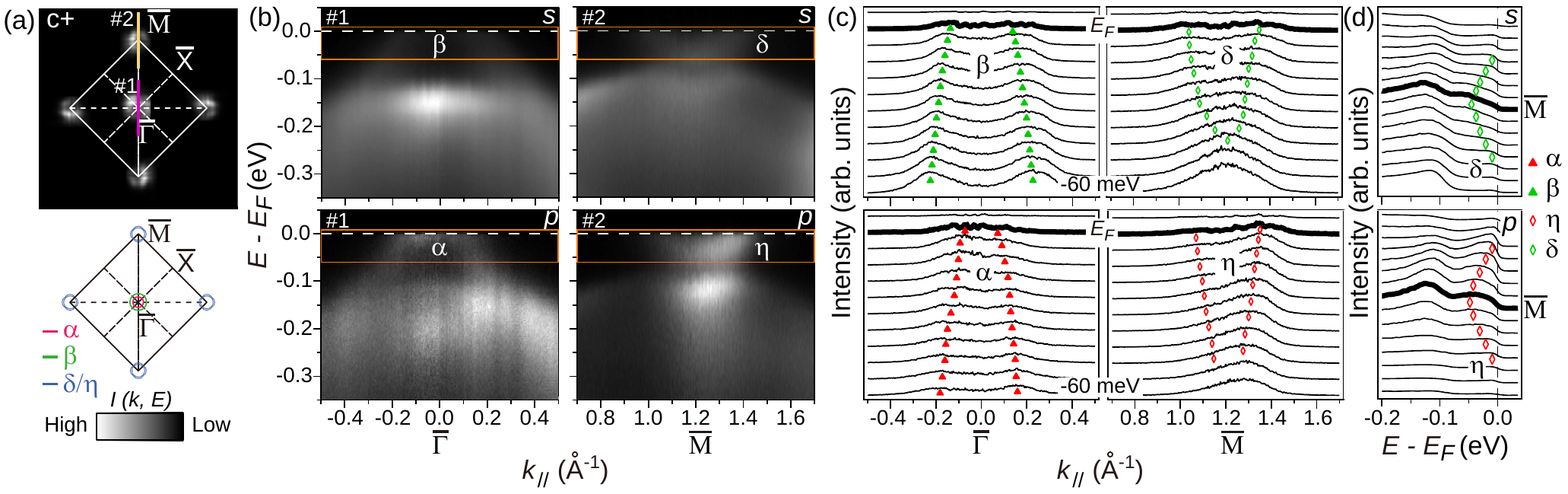}
\caption{(color online). (a) Photoemission intensity distribution (upper panel) integrated over [$E_F-10$~meV, $E_F$+10 meV] for an FeS single crystal taken with right circular polarized (C$+$) photons and the corresponding Fermi surface topology (lower panel). Here $\bar{\Gamma}$ and $\bar{M}$ stand for the two-dimensional projected Brillouin zone (BZ) center and corner, respectively. (b) Photoemission intensity distributions of the two cuts \#1 and \#2 identified by solid purple and yellow lines, respectively, in panel (a) taken with $s$ and $p$ polarized photons as indicated. (c) The momentum distribution curves (MDCs) in the corresponding rectangular areas of panel (b). (d) The energy distribution curves (EDCs) around the BZ corner (cut \#2) taken with $s$ and $p$ polarized photons as indicated.  The data around the BZ center and corner were taken with 24~eV and 30~eV photons, respectively, at SSRL.}
\end{figure*}

\begin{figure*}[htb]
\centering
\includegraphics[width=0.95\textwidth]{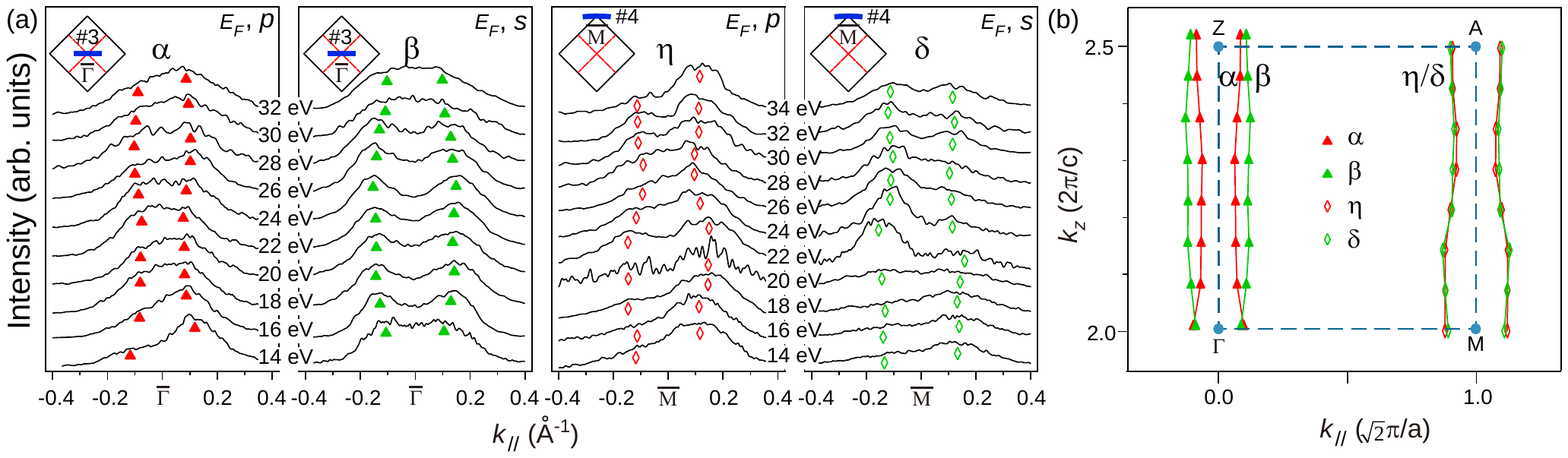}
\caption{(color online). (a) MDCs for the $\alpha$, $\beta$, $\eta$, and $\delta$ bands around $E_F$ as a function of photon energy, shifted for
clarity. The insets indicate the corresponding cuts in the BZ. They are distinguished by using either $s$ or $p$ polarization. (b) The experimental Fermi surface cross-section in the Z-$\Gamma$-M-A plane. Fermi surfaces (solid lines) are determined based on the experimental Fermi crossings (points). All data were taken at UVSOR.}

\end{figure*}

In this paper, we study the electronic structure of single-crystalline FeS by performing high-resolution angle-resolved photoemission spectroscopy (ARPES). There are two electron-like bands around the Brillouin zone (BZ) corner and two hole-like bands ($\alpha$ and $\beta$) around the BZ center. The Fermi surface exhibits modest $k_z$ variation. Moreover, we deduce that the absence of the third hole-like band with the $d_{xy}$ orbital character, which was discovered in many other iron-based compounds \cite{DLFengARPES4}, is likely due to strong  defect scattering. Compared with density functional theory (DFT) calculations\cite{SubediDFT}, the $\alpha$ and $\beta$   bands   near the Fermi energy ($E_F$)  are renormalized by factors of  3 and 2.5, respectively, indicating sizable electron correlations. By studying the relation between $T_c$ and the $\beta$ bandwidth of various  isovalently substituted FeSe compounds, we find that the overall evolution of $T_c$  is closely related to the strength of electron correlations.

\section{II. Experiment and Results}

So far, FeS single crystals could not be synthesized from elemental Fe and S as for FeTe and FeSe \cite{HuangTP}.  Instead, the
FeS single crystals used in our studies were synthesized by de-intercalation of K from K$_{0.8}$Fe$_{1.6}$S$_2$ precursor with a hydrothermal method \cite{LiTC2}. Since the starting material is K$_{0.8}$Fe$_{1.6}$S$_2$, whose Fe vacancy concentration is known to be quite high, the hydrothermal de-intercalation will likely make samples with Fe$_{1-x}$S$_{1-y}$. Here, the $x$ and $y$ should be quite  close, since the Fe:S ratio is almost 1:1 within the experimental error $\pm1$~$\%$ according to the chemical analysis \cite{LiTC2}.
Moreover, our samples exhibit quite high residual resistivity ratios (RRR) of about 40 \cite{SYLiHPTP}, which is better than the typical RRR of FeCh (Ch = Te, Se, S) and close to the one as reported in Refs.~\cite{SahanaTP, MahmoudTP2, JPSunHP}. Therefore,  the $x$ and $y$ should be quite  small. This is also consistent with   the sizable  $T_c$ of the sample,  since a few percent of defects at the Fe sites would normally kill the superconductivity.
Taken FeSe as an example, only 2~\% Cu doping on the Fe site would fully suppress the superconductivity \cite{TWHuangTP}. Therefore for simplicity, we still use the nominal chemical formula of FeS hereafter.
The photoemission data were collected with a Scienta R4000 electron  analyzer at Beamline 5-4 of the Stanford Synchrotron Radiation Laboratory (SSRL) and an MBS A-1 analyzer at Beamline 7U of the ultraviolet synchrotron orbital radiation facility (UVSOR). Both beamlines are equipped with an elliptically-polarized undulator which can switch the photon polarization among horizontal, vertical, and circular modes. The overall energy resolution is set to be 15 meV or better, and the typical angular resolution is 0.3~degrees. The samples were all cleaved \emph{in situ} and measured around 10~K under ultra-high vacuum better than 3$\times$10$^{-11}$ torr. Sample aging effects were carefully monitored to ensure they did not cause any artifacts in our analyses and conclusions.

The photoemission intensity distribution of FeS around $E_F$ is shown in Fig.~1(a). The spectral weight is mainly located around the BZ center and corner. To clearly resolve the band structure near $E_F$, we measured along the two cuts $\#1$ and $\#2$ crossing the BZ center and corner with both $s$- and $p$-polarized photons as shown in Fig.~1(b), and these two polarizations correspond to odd and even symmetries with respect to the mirror plane defined by the incoming light and outgoing photoelectron, respectively. According to the momentum distribution curves (MDCs) and the energy distribution curves (EDCs) in Figs.~1(c) and 1(d), two hole-like bands ($\alpha$ and $\beta$) are located around the BZ center, and two electron-like bands ($\eta$ and $\delta$) are located around the BZ corner. Based on the polarization of incident photons, $\alpha$ and $\eta$ exhibit even symmetry, while $\beta$ and $\delta$ exhibit odd symmetry. Based on the identified band dispersions, two circular hole Fermi pockets around the BZ center  are determined as shown in the lower panel of Fig.~1(a). Fermi crossings of the $\eta$ and $\delta$ bands are too close to be distinguished, but they would contribute to two nearly degenerate electron Fermi pockets  around the BZ corner that are mutually orthogonal, as predicted by the band calculations on FeS \cite{SubediDFT}.

Since the electronic structures of iron-chalcogenides usually contain three hole-like bands around the BZ center \cite{DLFengARPES4}, the so-called $\gamma$ band having $d_{xy}$ orbital symmetry is missing here. Among the five bands near $E_F$ in the iron-based superconductors,  the $\gamma$ band is by far the most sensitive to defects \cite{DLFengARPES4}, and its renormalization factor in FeSe is around 9 \cite{BorinsenkoARPES}, which is much higher than those of the other bands. With increased Te or Co concentrations, the $\gamma$ bands of  FeSe$_{1-x}$Te$_x$, NaFe$_{1-x}$Co$_x$As, and LiFe$_{1-x}$Co$_x$As all quickly broaden, indicating the $\gamma$ band becomes more incoherent, and even become featureless  in some cases  \cite{DLFengARPES4}. For sulfur-substituted FeSe, $\gamma$ `disappears' for sulfur content higher than 15~\% \cite{WatsonARPES}, {\sl i.e}, its intensity  submerges into the background, and
becomes undetectable.  Furthermore, the defects on  iron sites would cause even stronger scatterings than those on pnictogen/chalcogen sites.
As the hydrothermal de-intercalation process in the  synthesis might  retain some of the Fe vacancies in the FeS layer,  the `absence' of the $\gamma$ band is highly likely caused by the scattering of the defects in both Fe and S sites. They could severely  broaden the $\gamma$ band, so that no feature could be resolved.


A rather flat band around 170~meV below $E_F$ is observed around the BZ center taken with $s$-polarized photons in Fig.~1(b). This flat band in iron-based superconductors usually exhibits $d_{z^2}$ orbital character and should be observed with $p$-polarized photons \cite{DLFengARPES5}. The observation of this flat band here is due to the particular experimental geometry at the Beamline 5-4 end-station at SSRL, where $s$ polarized photons contain a polarization component along the $z$ axis \cite{DWShenARPES}.

\begin{figure}[t!]
\centering
\includegraphics[width=0.48\textwidth]{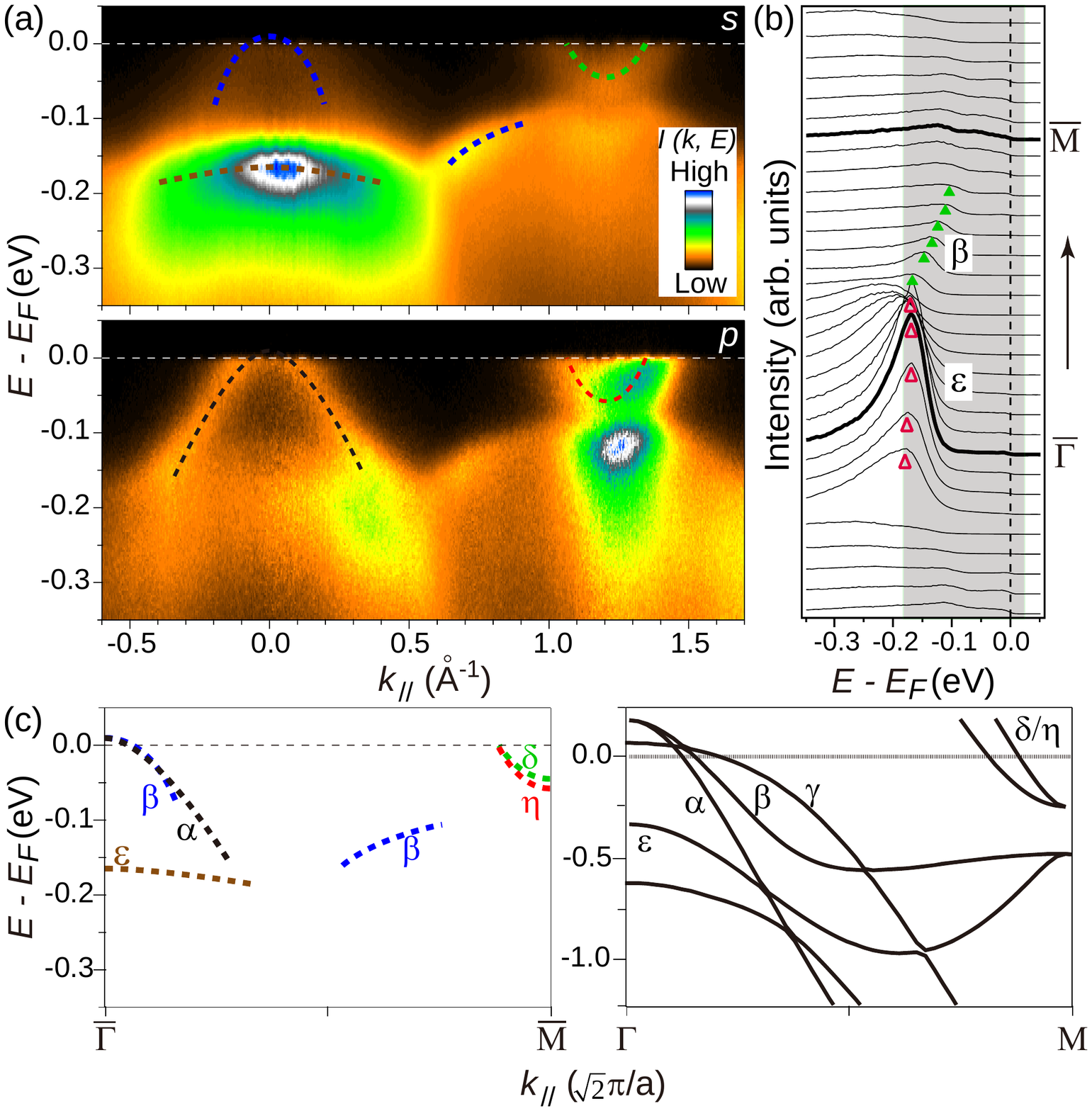}
\caption{(color online). (a) Photoemission intensity distributions along the $\Gamma$-M direction taken with $s$- and $p$-polarized photons,
respectively. (b) Corresponding energy distribution curves (EDCs) of data in panel (a) taken with $s$-polarized photons. (c) Left panel: the  band structure of FeS measured by ARPES. Right panel: the calculated FeS band structure along  $\Gamma$-M reproduced from Ref.~\cite{SubediDFT}. The calculation of the electronic structure was performed within the local-density approximation (LDA) with the general potential linearized augmented plane-wave (LAPW) method, including local orbitals. All the ARPES data were taken at SSRL with 30~eV photons.}
\end{figure}

\begin{figure*}[htb]
\centering
\includegraphics[width=0.95\textwidth]{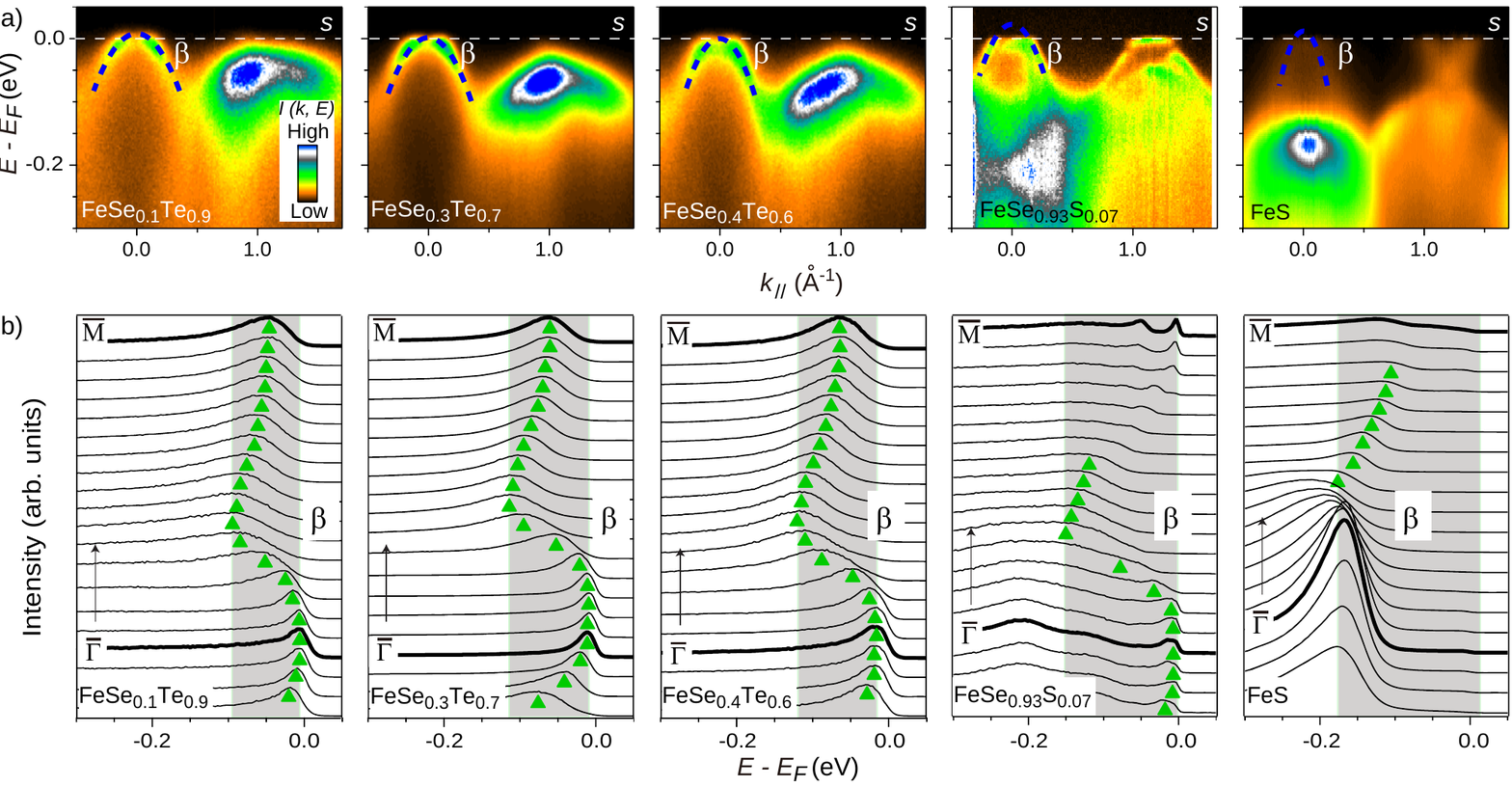}
\caption{(color online). (a) Photoemission intensity distributions along the $\Gamma$-M direction  taken with $s$-polarized photons for
FeSe$_{0.1}$Te$_{0.9}$, FeSe$_{0.3}$Te$_{0.7}$, FeSe$_{0.4}$Te$_{0.6}$, FeSe$_{0.93}$S$_{0.07}$, and FeS. (b) Corresponding EDCs of the data in panel (a). All the data except those of FeS were extracted from Refs.~\cite{DLFengARPES3, DLFengARPES4}.}
\end{figure*}

\begin{figure*}[htb]
  \begin{minipage}{0.5\textwidth}
    \includegraphics[width=0.95\textwidth]{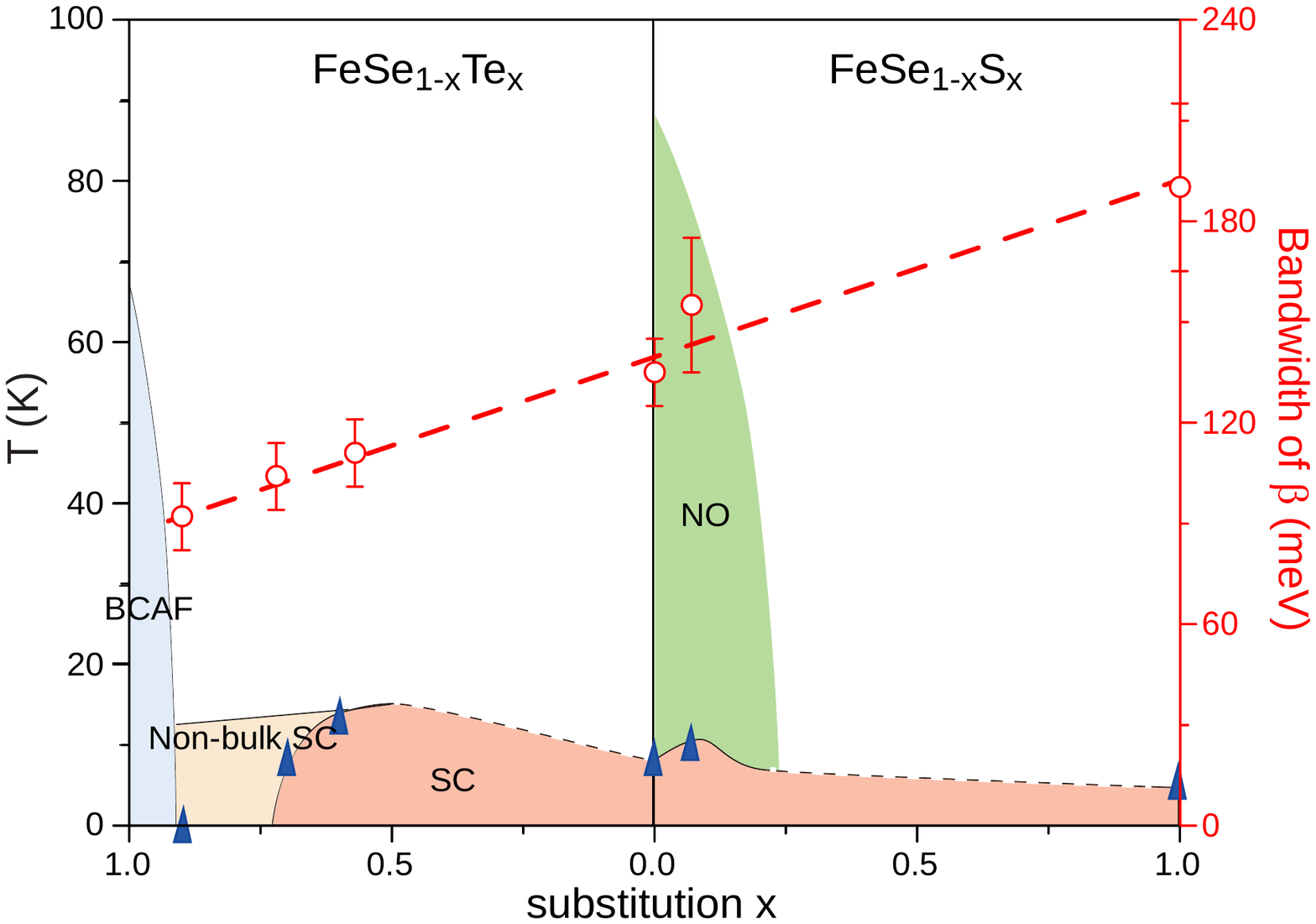}
  \end{minipage}%
  \begin{minipage}{0.5\textwidth}
    \caption{Phase diagram of $T_c$ and $\beta$ bandwidths as a function of the chalcogen content in both Te- and S-substituted FeSe. The superconducting, bicollinear antiferromagnetic, and nematic-ordered phases are     abbreviated as SC, BCAF, and NO, respectively. The bandwidths of FeSe$_{1-x}$Te$_x$ and FeSe$_{0.93}$S$_{0.07}$ were taken from Refs.~\cite{DLFengARPES3, DLFengARPES4}, and the phase diagram, including the $T_c$s for FeSe$_{1-x}$Te$_x$ and FeSe$_{1-x}$S$_x$ were taken from Refs.~\cite{WatsonARPES, MahmoudTP, LiuNS, PetrovicTP}. Two regions of the SC dome  marked by black dashed lines  are inferred based on interpolation, in the absence of any report on these substitution.}
  \end{minipage}
\end{figure*}

To illustrate the $k_z$ dependence of the Fermi crossings, Fig.~2(a) plots the MDCs along the $\Gamma$-M direction for these four bands near $E_F$, taken with both $s$- and $p$-polarized light and many different photon energies. The Fermi surface cross-section in the $\Gamma$-Z-A-M plane can be traced by the peak positions of the MDCs [Fig.~2(b)]. The Fermi momentum ($k_F$) of $\alpha$ first shrinks and then expands along the $\Gamma$-Z direction,  in contrast to that of the $\beta$ band. Meanwhile, the $k_F$s of both $\eta$ and $\delta$ shrink along the M-A direction, and they almost always coincide with each other within the experimental resolution. In general, the Fermi surface of FeS exhibits a quasi-two-dimensional behavior, consistent with quantum oscillation data  \cite{UjiQO}. Compared with BaFe$_2$(As$_{0.7}$P$_{0.3}$)$_2$ and Ba$_{1-x}$K$_x$Fe$_2$As$_2$, FeS is more two dimensional  \cite{DLFengARPES6}. Based on Luttinger's theorem, we estimate the electron concentration  to be 0.12 electrons per unit cell through the volume of the three-dimensional Fermi surface of the four detected bands \cite{Luttinger}. Since FeS should have electron-hole balanced Fermi surface volumes like FeSe \cite{ZXShenARPES2} and the compositions ratio of Fe : S here is almost 1:1, the electron and hole pockets should normally have the same total volume. Therefore, there should be a missing Fermi pocket with 0.12 holes, which can be naturally attributed to the $\gamma$ band.

As shown in Fig.~1(c), the dispersion of $\beta$ exhibits parabolic behavior. By fitting its dispersion below $E_F$ by a quadratic curve [Fig.~3(a)], we estimate the position of the $\beta$ band top to be  about 10~meV above $E_F$, and its effective band mass to be -1.74~$m_e$ ($m_e$ is the free electron mass). Meanwhile, the $\beta$ band bottom overlaps with the $\epsilon$ band as shown in both photoemission intensity distribution [Fig.~3(a)] and the EDCs [Fig.~3(b)]. The overlap occurs around 180~meV below $E_F$. If we take the full width at half maximum of the EDC peaks of $\epsilon$ as the error bar, and take the overlapping region as the $\beta$ band  bottom, the $\beta$ bandwidth is about 190$\pm$25~meV.  Similarly, we estimate the effective band mass of the $\alpha$ band is about -1.47~$m_e$.
 Fig.~3(c) reproduces  the band calculations  of FeS from Ref.\cite{SubediDFT}. Although  there are  subtle differences,
the observed band dispersions qualitatively agree with the calculations, for example,
$\alpha$ and $\beta$ are almost degenerate around the zone center in both the data and the calculations.
By comparing the  band masses obtained in both the data and the calculations, we obtain a renormalization factor of about 2.5 for the $\beta$ band, and 3 for $\alpha$, which indicate sizeble electronic correlations in FeS.

To position FeS in the bigger picture of the so-called  `11' series of bulk iron chalcogenide superconductors,
 photoemission intensity distributions of FeSe$_{1-x}$Te$_{x}$ and FeSe$_{1-x}$S$_{x}$ along the $\Gamma$-M direction and their corresponding EDCs for various substitutions are collected in Fig.~4.  All the data here were  taken with $s$-polarized photons, to emphasize $\beta$ band.
 Since the $\beta$ bandwidth can be easily estimated, it is taken as a characterization of the correlations in these materials.
Figure~4 shows that the bandwidth of $\beta$ increases monotonically from the tellurium end  in FeSe$_{1-x}$Te$_{x}$ to the sulfur end of  FeSe$_{1-x}$S$_{x}$, as expected from the decreasing bond length  \cite{SubediDFT}.  Consistent with the evolution of the bandwidth, the effective mass of $\beta$ near the BZ center, obtained by fitting the parabolic curves in Fig.~4(a), decreases monotonically from the tellurium end  in FeSe$_{1-x}$Te$_{x}$ to the sulfur end of  FeSe$_{1-x}$S$_{x}$.

In Fig.~5, the $\beta$ bandwidths are plotted onto the phase diagram of FeSe$_{1-x}$Te$_{x}$ and FeSe$_{1-x}$S$_{x}$. Although more data points are required to make a comprehensive case, the existing data fall on a line, as a function of substitution. From the FeTe end, bulk superconductivity emerges when  the bandwidth exceeds a certain value, and persists all the way to the FeS end. The $T_c$ of FeSe$_{1-x}$Te$_{x}$ and FeSe$_{1-x}$S$_{x}$ is enhanced at first, and then generally weakened with increased bandwidth. However, the relation between $T_c$ and substitution  is not monotonic. The superconductivity strengthens in the lightly S-substituted regime, likely due to the enhanced $(\pi,0)$ spin fluctuations related to the nematic order or to subtle Fermi surface topology effects
\cite{WatsonARPES}.

\section{III. Discussions}

It has been shown for almost all the iron-based superconductors that their phase diagrams can be understood from the bandwidth perspective \cite{DLFengARPES4,DLFengARPES7}. In particular, since isovalent substitution does not alter carrier density or the Fermi surface \cite{ZXShenARPES2, DLFengARPES7}, the strength of electron correlations, which can be represented by the inverse of bandwidth, would control the superconductivity of FeSe$_{1-x}$Te$_{x}$ and FeSe$_{1-x}$S$_{x}$. The  measured $\beta$ bandwidth of FeS is about 40\% larger than that of FeSe, which suggests  weaker correlation effects in FeS than in FeSe. On the other hand,   the sizable renormalization factors of FeS bands indicate that the electrons in FeS still experience significant interactions amongst themselves and with bosonic excitations, such as magnons and phonons. This might explain the robustness of superconductivity in FeSe$_{1-x}$S$_x$. This is reminiscent of Ba$_{1-x}$K$_x$Fe$_2$As$_2$, in which strong correlations exist throughout the entire doping range, allowing superconductivity to persist \cite{DLFengARPES4}.

It is also worthwhile to point out that  in both FeSe$_{1-x}$S$_{x}$ and K$_x$Fe$_2$Se$_{2-y}$S$_y$, $T_c$ decreases with S substitution, although these two families have different Fermi surface topologies.
The superconducting range  of $\beta$ bandwidth is 100$\sim$200 meV for Rb$_x$Fe$_2$Se$_{2-z}$Te$_z$  and  K$_x$Fe$_2$Se$_{2-y}$S$_y$ \cite{DLFengARPES7}. Superconductivity in FeSe$_{1-x}$Te$_{x}$ and FeSe$_{1-x}$S$_{x}$ also emerges for a bandwidth of $\sim$100~meV, but is  still  not suppressed at 190~meV (FeS end member).

Our data clearly imply that the superconductivity in FeS can survive without coherent quasiparticles in the $\gamma$ band. However, the incoherent spectral weight of $\gamma$ might  still contribute to the zero-energy excitations at low temperatures. This could explain  the nodal-gap-like behavior observed in thermal conductivity and specific heat  measurements  \cite{LiTC2, WenSHC}. On the other hand,  because the incoherent $\gamma$ band would not contribute to the superfluid response, if the superconducting gaps in the other bands are nodeless,  $\mu$SR could still observe an overall nodeless-gap behavior for FeS \cite{KirschnerMSR, HolensteinMSR}. Our results  may thus help resolve the contradictory reports by these techniques on the FeS gap structure.

\section{IV. Conclusions}

In summary, we have studied the electronic structure of superconducting FeS. Two hole-like bands and two electron-like bands around the BZ center and corner, respectively, have been resolved. The third hole-like band near $\Gamma$  (the $\gamma$ band)  may be too strongly scattered by defects to be observed, which may help resolve the current debate on the gap structure. The $k_z$ dispersions of the Fermi surfaces of FeS exhibit quasi-two-dimensional behavior, and the two electron-like bands are almost degenerate around $E_F$ within our resolution. Using the $\beta$ bandwidth as an indication of correlation strength, we illustrate the evolution of $T_c$ with electron correlation in FeSe$_{1-x}$Te$_{x}$ and FeSe$_{1-x}$S$_{x}$, and explain the robustness of superconductivity which still exists in the end member FeS. The observed electronic structure of FeS establishes a concrete foundation for further theoretical calculations and will help understand its superconducting properties.

\section{Acknowledgments}

The authors thank Dr.\ D.\ H.\ Lu for the experimental assistance at SSRL. This work is supported in part by the National Natural Science Foundation of China (Grant NO. 11604201, 11574194), National Key R\&D Program of the MOST of China (Grant No. 2016YFA0300203), Open Project Program of the State Key Lab of Surface Physics (Grant No.\ KF2016\_08), Fudan University, Science Challenge Program of China, and Shanghai Municipal Science and Technology Commission.

\end{document}